\documentclass[english,prd,nofootinbib,showpacs,preprint,floatfix,tightenlines]{revtex4}
\usepackage[T1]{fontenc}
\usepackage[latin1]{inputenc}
\usepackage{graphicx}

\makeatletter

\makeatletter

\usepackage{color}

\makeatletter

\makeatletter

\makeatletter

\usepackage{latexsym}

\usepackage{dcolumn}

\RequirePackage{graphicx}

\makeatother

\makeatother

\makeatother

\makeatother

\usepackage{babel}
\makeatother
\begin{document}

\preprint{arXiv:0810.0020 {[}hep-ph], ANL-HEP-PR-08-59, NSF-KITP-08-46, SMU-HEP-08-16}

\title{Longitudinal Parity-violating Asymmetry in Hadronic Decays of Weak
Bosons in Polarized Proton Collisions}

\author{Edmond~L.~Berger}

\affiliation{High Energy Physics Division, Argonne National Laboratory, Argonne,
Illinois 60439, USA}

\author{Pavel~M.~Nadolsky}

\affiliation{Department of Physics, Southern Methodist University, Dallas, TX
75275, USA}

\date{\today}

\begin{abstract}
We investigate the possible measurement of parity-violating spin asymmetries
in jet pair production in proton-proton collisions at the Brookhaven
Relativistic Heavy Ion Collider (RHIC), with the goal to constrain
longitudinally polarized quark and antiquark distribution functions.
A measurable asymmetry could be observed in the vicinity of the massive
weak $W$ boson resonance, where the parity-violating signal appears
above the parity-conserving background and is enhanced by interference
of the strong and electroweak production amplitudes. We discuss the
potential for such measurements, perhaps the first opportunity to
measure a parity-violating asymmetry in proton-proton collisions at
RHIC. Sensitivity of this measurement to the polarization of down-type
antiquarks is demonstrated. 
\end{abstract}

\pacs{12.38.Bx, 13.87.-a, 13.88.+e}

\maketitle

\section{Introduction}

\label{sec:introduction} The asymmetry in the rapidity-dependent
cross section for weak ($W$) boson production in unpolarized high-energy
$p\overline{p}$ collisions is a valuable instrument~\cite{Berger:1988tu}
for measuring the Bjorken $x$ dependence of the ratio $u(x,M_{W})/d(x,M_{W})$
of the up-quark and down-quark quark parton distribution functions
(PDFs) in a proton at the very hard scale set by the weak boson mass
$M_{W}$. Demonstration of the utility of this prediction awaited
sufficient luminosity \cite{Abe:1991cd,Abe:1994rj}, but it was not
long before the measured rapidity asymmetries were used routinely
in global analyses as important constraints on unpolarized PDFs \cite{Martin:1992as,Lai:1996mg}.
The method relies on the accepted Drell-Yan mechanism by which the
$W^{\pm}$ boson is produced in the standard model, e.g., principally
$u+\bar{d}\rightarrow W^{+}$ and $d+\bar{u}\rightarrow W^{-}$ at
leading order in quantum chromodynamics (QCD) perturbation theory.
Thus, at least qualitatively, a $W^{+}$ moving with a large rapidity
$y$ is produced from projectile $u$ quarks carrying large $x$,
whereas a $W^{-}$ moving with a large $y$ is produced from projectile
$d$ quarks carrying large $x$. The ratio of the unpolarized $W^{+}$
and $W^{-}$ cross sections at large $y$ thus provides important
constraints on the ratio $u(x,M_{W})/d(x,M_{W})$ at $x$ between
0.1 and 1.

Extension of similar arguments to polarized $pp$ scattering~\cite{Berger:1988unpub,Bourrely:1993dd,Bourrely:1994sc,Nadolsky:1995nf,Gehrmann:1997ez,Gluck:2000ek}
adds a promise of direct access to the longitudinal spin-dependent
quark PDFs $\Delta q(x,M_{W})$ via the $(V-A)$ coupling of the quark
constituents to the $W$. Only one of the incident protons need be
longitudinally polarized in order for this parity-violating measurement
to be effective. In this paper we address $W$ production in longitudinally
polarized proton collisions at the Brookhaven Relativistic Heavy Ion
Collider (RHIC) at collision energy $\sqrt{s}=500$ GeV~\cite{Bunce:2000uv}.
We focus on the dijet decay mode of the $W$, $pp\rightarrow(W\rightarrow jj)+X$,
and compute fully differential cross sections in jet pair invariant
mass $Q$, jet pair rapidity $y$, and jet transverse momentum $p_{Tj}$
for all signal and background processes.

The parity-violating dijet production cross section, $\Delta\sigma=\sigma(p^{\rightarrow}p)-\sigma(p^{\leftarrow}p),$
where $p^{\rightarrow}$ ($p^{\leftarrow}$) represents a proton with
its spin aligned with (against) the proton's direction of motion,
is dominated by the $u\overline{d}+\overline{d}u\rightarrow W^{+}$
process. The single-spin asymmetry \begin{equation}
A_{L}=\frac{\sigma(p^{\rightarrow}p)-\sigma(p^{\leftarrow}p)}{\sigma(p^{\rightarrow}p)+\sigma(p^{\leftarrow}p)},\label{AL}\end{equation}
 (as a function of the rapidity $y$ ) is sensitive to $\Delta u(x,M_{W})$
at $x\rightarrow1$ at large positive $y$ values, and to $\Delta\overline{d}(x,M_{W})$
at $x\rightarrow M_{W}^{2}/s$ at large negative $y$ values, as we
demonstrate quantitatively. Both $\Delta u$ and $\Delta\overline{d}$
PDFs are not known well in the quoted $x$ regions, as they are not
constrained by the existing data (mostly from polarized deep inelastic
scattering). The dijet asymmetry $A_{L}$ may probe these PDFs with
a relatively small event sample.

Commonly one detects a $W$ boson by observing a leptonic decay $W\rightarrow\ell+\nu$,
with its characteristic Jacobian peak in the transverse momentum $p_{T}$
of the charged leptons $\ell=e$ or $\mu$, and the missing transverse
energy carried by the invisible neutrino $\nu$.%
\footnote{For a general discussion of production and observation of $W$ bosons
at hadron colliders, see, \emph{e.g.,} Ref.~\cite{BargerPhillips1988}. %
} The lepton decay mode is observable at all colliders, including RHIC
\cite{Nadolsky:2003ga,Nadolsky:2003fz,Kamal:1997fg}. Such signals
are clean and impose tight constraints on the PDFs, but the small
decay branching fraction penalty is felt keenly in the event rate.
The luminosity in polarized proton scattering tends to be less than
in the unpolarized case, and the $W$ production cross section at
the envisioned RHIC center-of-mass energy is not as great as at a
higher-energy collider. Correspondingly, lepton decay events come
at a premium and most likely would require several years of RHIC running.
With this in mind, it is attractive to investigate the {\em a priori}
disfavored alternative that one could observe $W$ production in the
hadronic decay mode $W\rightarrow{\rm 2\mbox{ jets}}$ \cite{Bourrely:1990pz},
precisely the topic of this paper.

An advantage of the hadronic $W$ decay compared to the leptonic decay
is a larger cross section, enhanced by $\mbox{Br}(W\rightarrow\mbox{hadrons})/\mbox{Br}(W\rightarrow\ell\nu)\approx6$.
Another advantage is the possibility to directly determine the invariant
mass and rapidity of the $W$ boson, approximately equal in this case
to the invariant mass ($Q$) and rapidity ($y$) of the high-$p_{T}$
jet pair. The boson-level $y$ distribution does not suffer from the
smearing of the PDF dependence in the charged-lepton distributions
\cite{Nadolsky:2003ga} caused by spin correlations in the course
of the $W$ boson decay.

The hadronic decay mode of the $W$ boson has an admitted drawback
in that background contributions to jet pair production are substantial.
Hadronic jets are produced most abundantly through QCD and electromagnetic
(virtual photon) interactions, rather than through $W$ interactions,
especially via the scattering involving real gluons, which grows rapidly
with collider energy. The backgrounds present a major obstacle for
observing unpolarized electroweak ($W$ and $Z$ mediated) jet production
at the Fermilab Tevatron $p\bar{p}$ collider and the CERN Large Hadron
Collider \cite{Pumplin:1991fy,Baur:2000xd}. At the intermediate energy
of RHIC, the background/signal ratio (25-30 or more) is comparable
to that at the CERN $p\overline{p}$ Super Proton Synchrotron (SPS)
with the energy $\sqrt{s}=630$ GeV, where a $3\sigma$ enhancement
from $W$ and $Z$ bosons in the dijet mass ($Q$) distribution was
observed \cite{Ansari:1987vf} with very low integrated luminosity
(${\cal L}=0.73\mbox{ pb}^{-1}$).

We argue that the main backgrounds can potentially be controlled at
RHIC because of their smooth (non-resonant) kinematical dependence
and symmetry with respect to spatial reflections. Due to the smooth
behavior, side-band subtraction can be used to advantage to estimate
the \emph{unpolarized} background accurately in the $W$ signal region
by extrapolation from the side bands of the $Q$ distribution. The
parity symmetry of the dominant background leads to its cancellation
in the single-spin cross section, up to a small false asymmetry due
to the uncertainty in the proton beam polarization. Thus the parity-violating
$A_{L}$ preserves the clear pattern of PDF flavor dependence typical
of $W$ boson production and allows us access to $\Delta u(x,M_{W})$
and to $\Delta\overline{d}(x,M_{W})$ in well-defined kinematic regions.

It is not possible to distinguish $W^{+}$ and $W^{-}$ contributions
in the jet pair decay mode. This apparent limitation does not disfavor
the hadron decay mode with respect to the leptonic decay mode in situations
in which charged lepton identification is not available, such as at
the earlier stages of RHIC spin program. Full separation of the $W$
and $Z$ signals is not possible as a consequence of limited jet invariant
mass resolution (of order 5-10 GeV). However, the $Z$ contribution,
with its relatively small event rate, has only mild effect on the
PDF dependence of jet pair cross sections.

We evaluate all scattering processes for production of jet pairs (approximated
by a pair of final-state partons) at the leading order in the QCD
and electroweak coupling strengths, $\alpha_{s}$ and $\alpha_{EW}$.
Spin-dependent distributions in dijet mass ($Q$) and jet transverse
momentum ($p_{Tj}$) at this order are reported in Ref.~\cite{Bourrely:1990pz}.
In our paper, we focus on fully differential distributions and comparison
of dijet production channels, issues essential for the observation
of $A_{L}.$ We compute these distributions using a modified Monte-Carlo
integrator MadEvent \cite{Stelzer:1994ta,Maltoni:2002qb,Alwall:2007st}.
The presence of both strong and electroweak contributions to the same
final state means that their constructive and destructive interference
is possible. We compute the magnitude of the interference effect in
both the unpolarized and the single-spin cross sections. Although
the interference effect enhances the single-spin cross section considerably,
it does not alter the sensitivity of $A_{L}$ to the polarized PDFs.

In a recent paper \cite{Arnold:2008zx}, Arnold {\em et al.} also
consider jet production in polarized proton proton collisions at RHIC
energies. Our focus and emphasis are somewhat different from theirs.
In our work, we compute fully differential cross sections of two jets,
whereas Ref.~\cite{Arnold:2008zx} investigates the single-jet inclusive
double-differential cross section $d\sigma/(dp_{Tj}dy_{j})$, integrated
over the full range of dijet mass $Q$. All background contributions
are included in the unpolarized denominator of $A_{L}$ in Ref.~\cite{Arnold:2008zx},
reducing predicted asymmetry values to about 1\% or less. In our study,
we find that selection cuts on $Q$ and other kinematical variables,
combined with the subtraction of the background in the signal region
$70<Q<90$ GeV, are very effective in suppressing unpolarized dijet
production. Owing to the use of these techniques our predictions for
spin asymmetries are larger at least by an order of magnitude, and
our appraisal of the jet pair mode is more optimistic.

A tag on final-state charmed hadrons is proposed in Ref.~\cite{Arnold:2008zx}
as a means to improve the signal over the background. In principle
charm tagging eliminates the dominant gluon and light-quark parity-conserving
processes, while preserving a large fraction {[}of order $\mbox{Br}\left(W\rightarrow cX\right)/\mbox{Br}\left(W\rightarrow\mbox{all}\right)\approx33\%$]
of electroweak signal events. In practice, the utility of this approach
is diminished by the low efficiency of the experimental identification
of final-state charmed hadrons. Charm scattering contributions are
included in our Monte-Carlo calculation, so that charm tagging can
be explored together with the other handles at the fully differential
level.

In Sec.\ \ref{sec:Unpolarized}, we discuss the basic QCD processes
that produce a pair of jets. We show figures that illustrate the size
of various QCD and electroweak contributions to the cross section,
as a function of the dijet invariant mass and other variables. To
obtain our quantitative estimates of polarization asymmetries in Sec.\ \ref{sec:Asymmetry},
we must adopt parametrizations of spin dependent parton distributions.
We choose to work with the de Florian-Navarro-Sassot parametrization~\cite{deFlorian:2005mw}.
Using these PDFs, we explore sensitivity of the predictions for $A_{L}$
to variations in the spin-dependent PDF parametrizations.

Conclusions are summarized in Sec.\ \ref{sec:Conclusions}. Our quantitative
predictions show that the predicted magnitude of $A_{L}$ is sufficiently
large relative to the anticipated uncertainties, so that significant
measurements should be possible, with good prospects for discrimination
among different polarized PDFs. We urge experimental study of $A_{L}(y)$
in hadronic decays of $W$ at RHIC.

\section{Unpolarized jet pair production \label{sec:Unpolarized}}

\subsection{General remarks \label{sub:General-remarks}}

We consider inclusive jet pair production in proton-proton collisions
at $\sqrt{s}=500$ GeV, $pp\rightarrow jjX$, approximated by $2\rightarrow2$
exchanges of the standard model bosons ($V=g$, $\gamma^{*}$, $W^{\pm}$,
and $Z^{0}$) in the $s$, $t$, and $u$ channels, including the
interference between the amplitudes with different types of the bosons.
The unpolarized and single-spin cross sections, $\sigma$ and $\Delta\sigma$,
for production of an energetic jet pair are given by \begin{equation}
\frac{d\sigma}{d\vec{p}_{1}d\vec{p}_{2}}=\sum_{a,b,c,d}\frac{d\widehat{\sigma}(ab\rightarrow cd)}{d\vec{p}_{1}d\vec{p}_{2}}f_{a/p}(x_{1},Q)f_{b/p}(x_{2},Q),\end{equation}
 and \begin{equation}
\frac{d\Delta\sigma}{d\vec{p}_{1}d\vec{p}_{2}}=\sum_{a,b,c,d}\frac{d\Delta\widehat{\sigma}(ab\rightarrow cd)}{d\vec{p}_{1}d\vec{p}_{2}}\Delta f_{a/p}(x_{1},Q)f_{b/p}(x_{2},Q),\end{equation}
where $p_{1}^{\mu}$ and $p_{2}^{\mu}$ are the momenta of the jets,
$Q^{2}=(p_{1}+p_{2})^{2}$, $x_{1,2}=Qe^{\pm y}/\sqrt{s}$, $y=\ln\left[\left(Q^{0}+Q^{3}\right)/\left(Q^{0}-Q^{3}\right)\right]/2$,
$(\Delta)\widehat{\sigma}(ab\rightarrow cd)$ is the perturbative
cross section in the $ab\rightarrow cd$ parton scattering channel,
and $f_{a/p}(x,Q)$ ($\Delta f_{a/p}(x,Q)$) is the unpolarized (longitudinally
polarized) PDF. The QCD renormalization and factorization scales are
set equal to the invariant mass $Q$ of the dijet. The $z$ axis follows
the direction of the polarized proton beam. In some cases, we use
the notation $\sigma(V)$ for the partial cross section that involves
the same boson species $V$ in the scattering amplitude and its conjugate,
without the interference with other types of bosons.

The lowest-order cross sections for $s$-channel $W^{+}$ ($W^{-}$)
production via $q_{i}\overline{q}_{j}\rightarrow W^{\pm}\rightarrow q_{k}\overline{q}_{l}$
(the dominant parity-violating contribution) are \begin{eqnarray}
\frac{d\sigma(W,\, s\mbox{-channel})}{dQ^{2}dy} & = & \sum_{i,j}C_{ij}\left[q_{i}(x_{1},Q)\overline{q}_{j}(x_{2},Q)+\overline{q}_{j}(x_{1},Q)q_{i}(x_{2},Q)\right]\end{eqnarray}
 and\begin{eqnarray}
\frac{d\Delta\sigma(W,\, s\mbox{-channel})}{dQ^{2}dy} & = & \sum_{i,j}C_{ij}\left[-\Delta q_{i}(x_{1},Q)\overline{q}_{j}(x_{2},Q)+\Delta\overline{q}_{j}(x_{1},Q)\overline{q}_{i}(x_{2},Q)\right],\end{eqnarray}
 where\[
C_{ij}=\frac{\pi^{2}\alpha_{EW}^{2}}{12s}\frac{Q^{2}}{\left(Q^{2}-M_{W}^{2}\right)^{2}+\Gamma_{W}^{2}Q^{4}/M_{W}^{2}}\left|V_{ij}\right|^{2}\sum_{k,l}\left|V_{kl}\right|^{2},\]
 $\alpha_{EW}=g_{w}^{2}/4\pi$ is the electroweak coupling strength,
$M_{W}$ and $\Gamma_{W}$ are the $W$ boson mass and width, and
$V_{ij}=V_{ji}$ is the quark mixing (Cabibbo-Kobayshi-Maskawa) matrix.
The indices $i,$ $j,$ $k,$ $l$ are summed over flavors of quark-antiquark
pairs carrying the net electric charge of the $W^{+}$ or $W^{-}$
boson. At the energy of RHIC, where contributions with initial-state
$s,$ $c,$ $b$ (anti-)quarks are much smaller than those with $u$
and $d$ (anti-)quarks, the $W^{+}$ and $W^{-}$ cross sections are
well approximated by \begin{equation}
\frac{d\sigma(W^{+},\, s\mbox{-channel})}{dQ^{2}dy}=C_{ud}\left[u(x_{1},Q)\bar{d}(x_{2},Q)+\bar{d}(x_{1},Q)u(x_{2},Q)\right]+\mbox{ small terms};\label{sWplus}\end{equation}
 \begin{equation}
\frac{d\sigma(W^{-},\, s\mbox{-channel})}{dQ^{2}dy}=C_{ud}\left[d(x_{1},Q)\bar{u}(x_{2},Q)+\bar{u}(x_{1},Q)d(x_{2},Q)\right]+\mbox{ small terms};\label{sWminus}\end{equation}
 \begin{equation}
\frac{d\Delta\sigma(W^{+},\, s\mbox{-channel})}{dQ^{2}dy}=C_{ud}\left[-\Delta u(x_{1},Q)\bar{d}(x_{2},Q)+\Delta\bar{d}(x_{1},Q)u(x_{2},Q)\right]+\mbox{ small terms};\label{DsWplus}\end{equation}
 and \begin{equation}
\frac{d\Delta\sigma(W^{-},\, s\mbox{-channel})}{dQ^{2}dy}=C_{ud}\left[-\Delta d(x_{1},Q)\bar{u}(x_{2},Q)+\Delta\bar{u}(x_{1},Q)d(x_{2},Q)\right]+\mbox{ small terms}.\label{DsWminus}\end{equation}
 These resonant cross sections are combined with non-resonant ($t-$
and $u$-channel) contributions to form $\sigma(W)$, the full $W$-boson
contribution of order $\alpha_{EW}^{2}$. In a similar manner, we
construct pure $Z$ and $\gamma^{*}$ contributions, also of order
$\alpha_{EW}^{2}$. In addition, one must include the cross section
for the interference between the electroweak bosons (of order $\alpha_{EW}^{2}$),
and between the electroweak bosons and gluons (of order $\alpha_{S}\alpha_{EW}$).
Finally, there is a pure QCD cross section, of order $\alpha_{s}^{2}$.
Explicit matrix elements for all these processes can be found in Ref.~\cite{Bourrely:1990pz}.

Our fully differential spin-dependent cross sections for $2\rightarrow2$
jet production are computed in all channels with the help of the programs
MadGraph and MadEvent \cite{Stelzer:1994ta,Maltoni:2002qb,Alwall:2007st}.
MadGraph is a program for automatic generation of tree-level cross
sections in the standard model and its common extensions. MadEvent
realizes phase-space integration of these cross sections. Internally
these programs operate with helicity-dependent scattering amplitudes
obtained using the HELAS library \cite{Murayama:1992gi}. In a typical
setting, the amplitudes are summed over all helicity combinations
to produce spin-averaged cross sections. We modified {\nolinebreak
MadEvent} to also allow evaluation of single-spin cross sections
for an arbitrary scattering process, including jet pair production.%
\footnote{The modified MadEvent program for the calculation of polarized cross
sections is available from one of the authors (P.M.N.) by request.%
} This modified program is employed to evaluate both the numerator
and denominator of $A_{L}$.

In hard-scattering cross sections, we include contributions from four
quark flavors ($u,$ $d,$ $s,$ and $c$), in accordance with the
default MadGraph choice. The unpolarized and longitudinally polarized
PDFs are taken from the Martin-Roberts-Stirling-Thorne (MRST'2002
NLO \cite{Martin:2002aw}) and de Florian-Navarro-Sassot (DNS'2005
\cite{deFlorian:2005mw}) sets, respectively. The MRST'2002 NLO PDFs
are chosen because they satisfy positivity conditions with the DNS'2005
PDFs.

\begin{figure}
\begin{centering}\includegraphics[width=0.5\columnwidth,keepaspectratio]{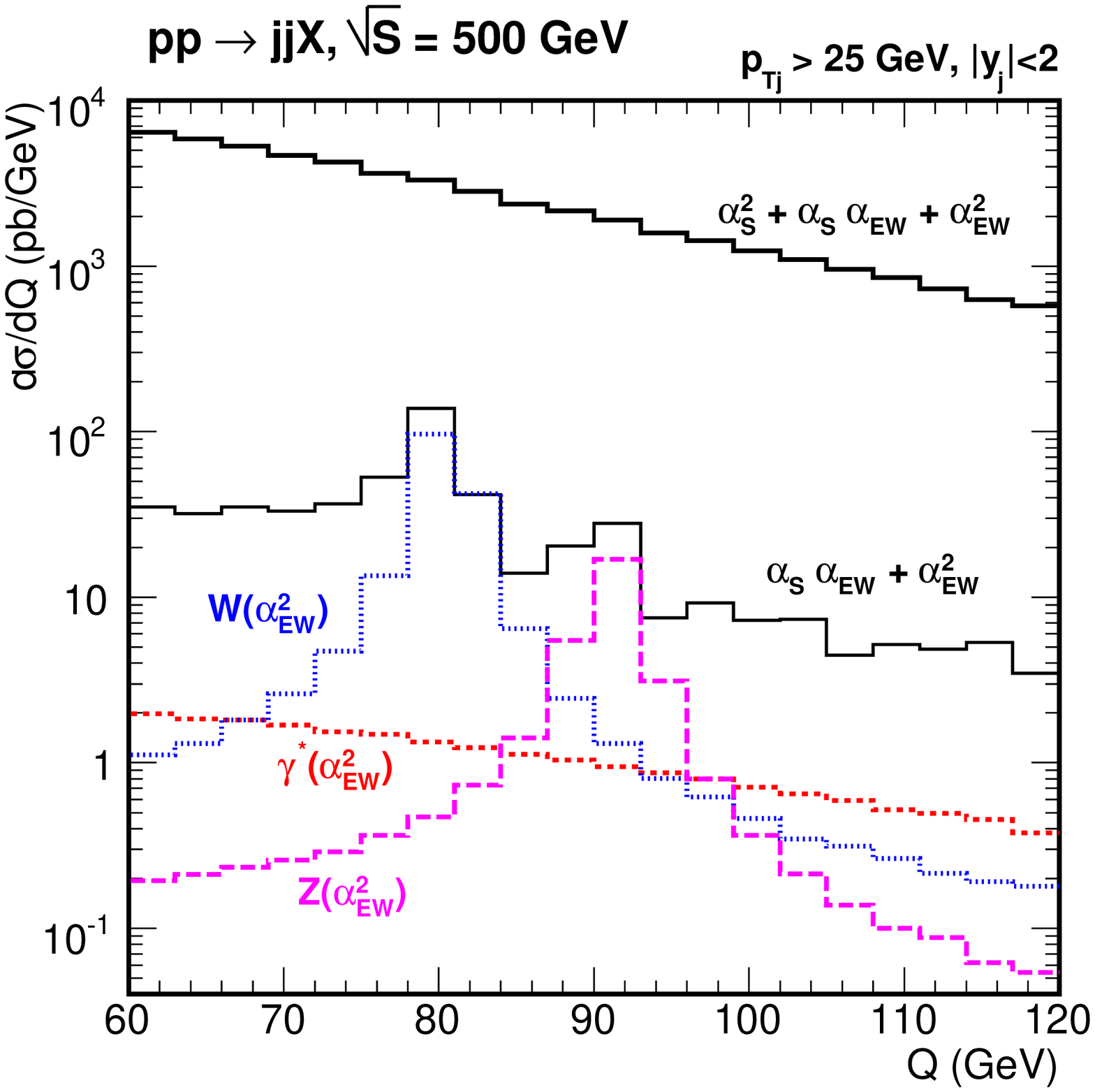}\includegraphics[width=0.5\columnwidth,keepaspectratio]{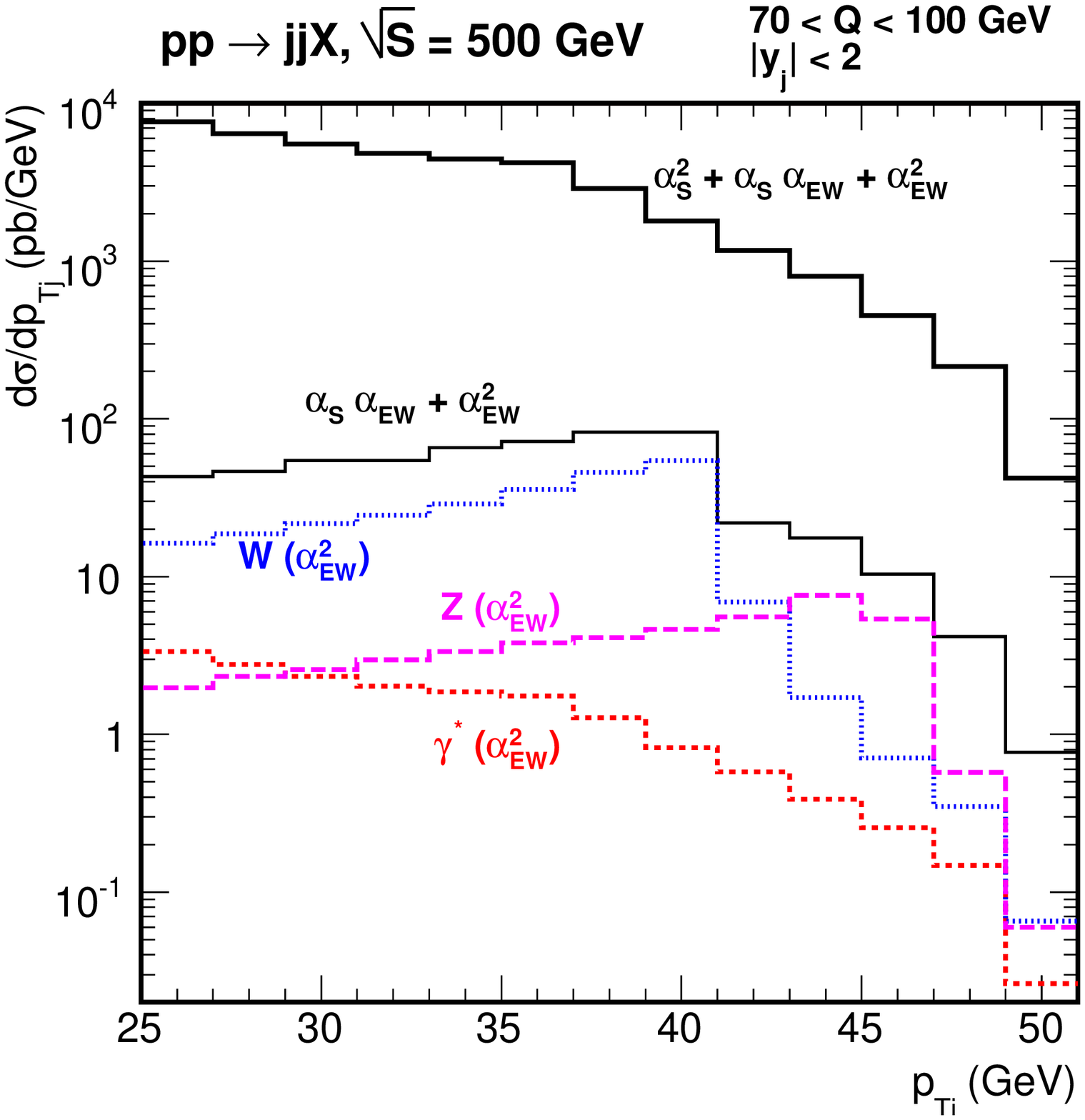}\\
 (a)\hspace{2.5in}(b)\par\end{centering}

\begin{centering}\includegraphics[width=0.6\columnwidth,keepaspectratio]{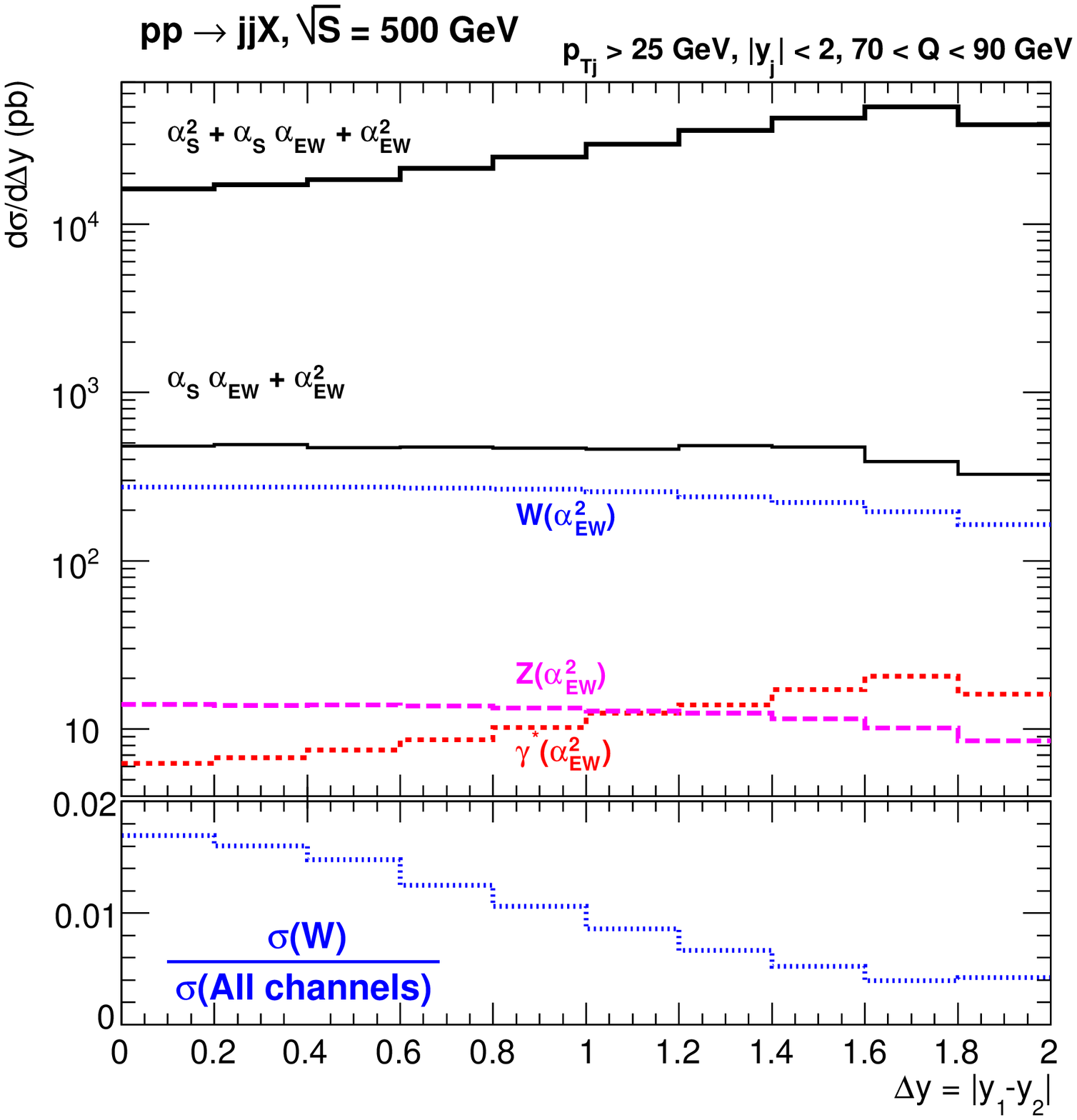}\\
 (c)\par\end{centering}

\caption{Unpolarized cross sections for $pp\rightarrow jjX$ at $\sqrt{s}=500$
GeV in various scattering channels, plotted as a function of (a) dijet
invariant mass $Q$, (b) jet transverse momentum $p_{Tj}$, and (c)
difference of jet rapidities $\Delta y=\left|y_{1}-y_{2}\right|$,
for the cuts specified in the figures. \label{Fig:unp}}
\end{figure}

\subsection{Kinematic distributions of unpolarized cross sections\label{sub:Kinematical-distributions}}

Spin-averaged jet production is dominated by the continuous event
distribution from QCD and electromagnetic scattering (involving only
$g$ and $\gamma^{*}$). These involve, in order of their magnitudes
at $60<Q<100$ GeV, $qg\rightarrow qg,$ $gg\rightarrow gg,$ $qq^{\prime}\rightarrow qq^{\prime},$
$qq\rightarrow qq,$ and smaller scattering contributions, where $q$
stands for both quarks and antiquarks, and $qq$ ($qq^{\prime}$)
stands for scattering of the same (different) quark flavors. The two
largest jet cross sections produced by $qg$ and $gg$ scattering
cancel when a cross section difference is taken to compute the single-spin
asymmetry. Parity violation needed to obtain a non-zero numerator
of $A_{L}$ in Eq.~(\ref{AL}) arises solely from $qq$ contributions
with intermediate $W$ and $Z$ bosons. The $s$-channel parity-violating
scattering amplitudes are enhanced resonantly when the dijet invariant
mass $Q$ is close to $M_{W}$ ($M_{Z}$), in the $Q$ range we will
call {}``the signal region''. Even in this region, the spin-averaged
$W$ and $Z$ contribution constitutes at most a few percent of the
full event rate, suggesting that the most straightforward measurement
(not separating the signal and background contributions) would result
in small spin asymmetries. The magnitude of $A_{L}$ may be enhanced
by applying a {}``side-band subtraction'' technique, i.e., by measuring
the large parity-conserving background at values of $Q$ outside of
the signal region and subtracting it from the denominator of $A_{L}$
inside the signal region.

Let us now turn to the figures illustrating these observations. The
spin-averaged differential cross sections for various combinations
of scattering channels are shown in Fig.~\ref{Fig:unp}. We focus
on the distributions in the dijet invariant mass $Q$, the transverse
momentum $p_{Tj}$ of the jet, and the difference $\Delta y$ of jet
rapidities, $\Delta y=\left|y_{1}-y_{2}\right|$, shown in Figs.~\ref{Fig:unp}(a,b,c).
Cross sections denoted {}``$W(\alpha_{EW}^{2})$'', {}``$Z(\alpha_{EW}^{2})$'',
and {}``$\gamma^{*}(\alpha_{EW}^{2})$'' represent the pure $W$
boson contribution, the $Z$ boson contribution, and the $\gamma^{*}$
contribution, without the interference terms. The sum of all electroweak
and interference cross sections is shown by a lower solid line labeled
{}``$\alpha_{s}\alpha_{EW}+\alpha_{EW}^{2}$''. The full cross section,
which also includes the pure QCD contribution of order $\alpha_{s}^{2}$,
is shown by the upper solid line with the label {}``$\alpha_{S}^{2}+\alpha_{s}\alpha_{EW}+\alpha_{EW}^{2}$''.

In most figures, we impose constraints $p_{Tj}>25$ GeV and $\left|y_{j}\right|<2$
to reproduce approximately the acceptance of the STAR detector~\cite{Harris:1993ck}.%
\footnote{Our results for an asymmetric cut $-2\leq y_{j}\leq1$ on jet rapidities
(exactly corresponding to the STAR acceptance) are qualitatively similar. %
} As seen in Figs.~\ref{Fig:unp}(a) and (b), $W$ ($Z$) production
receives a resonant enhancement at $Q\approx M_{W}$ ($M_{Z}$) and
$p_{Tj}\approx M_{W}/2$ ($M_{Z}/2)$. The QCD and electromagnetic
backgrounds fall smoothly with both $Q$ and $p_{Tj}.$ To focus on
the parity-violating asymmetry, we impose a selection $70<Q<Q_{max},$
with $Q_{max}=90-100$ GeV, in Figs.~\ref{Fig:unp}(b) and (c). A
value of $Q_{max}$ below $M_{Z}$ is generally preferred in order
to suppress the $Z$ cross section (constituting about 1/6 of $\sigma(W)$
at their respective mass poles according to Fig.~\ref{Fig:unp}(a))
and emphasize the PDF dependence typical for the $W$ boson contribution.

In $2\rightarrow2$ scattering, $\Delta y$ is related to the scattering
angle $\theta_{*}$ in the jet pair rest frame as $\tanh(\Delta y/2)=\cos\theta_{*}$.
Signal and background processes are characterized by different spin
correlations between the initial- and final-state particles and, therefore,
different shapes of $d\sigma/d(\cos\theta_{*})$ {[}or $d\sigma/d\Delta y$].
The kinematic differences between the signal and background $d\sigma/d\Delta y$
distributions are traced largely to real-gluon emissions present in
the QCD background process, but not in the electroweak processes (see
below). Figure~\ref{Fig:unp}(c) shows that the full ${\cal O}(\alpha_{S}^{2}+\alpha_{s}\alpha_{EW}+\alpha_{EW}^{2})$
cross section is peaked strongly at large $\left|\Delta y\right|,$
while the signal ${\cal O}(\alpha_{EW}^{2})$ and ${\cal O}(\alpha_{EW}^{2}+\alpha_{s}\alpha_{EW})$
cross sections have flatter $\left|\Delta y\right|$ dependence.In
the dijet rest frame, the ${\cal O}(\alpha_{S}^{2}+\alpha_{s}\alpha_{EW}+\alpha_{EW}^{2})$
cross section peaks more strongly at $\cos\theta_{*}\rightarrow\pm1$
as compared to the ${\cal O}(\alpha_{EW}^{2}+\alpha_{s}\alpha_{EW})$
cross section.

The full electroweak+interference cross section constitutes 2-3\%
of the full cross section at $\left|\Delta y\right|=0$, but it drops
below 1\% at $\left|\Delta y\right|\approx2.$ This difference could
be exploited to enhance the signal/background ratio. For simplicity,
we don't impose this selection in the rest of the paper.

\begin{figure}
\includegraphics[width=0.5\columnwidth,keepaspectratio]{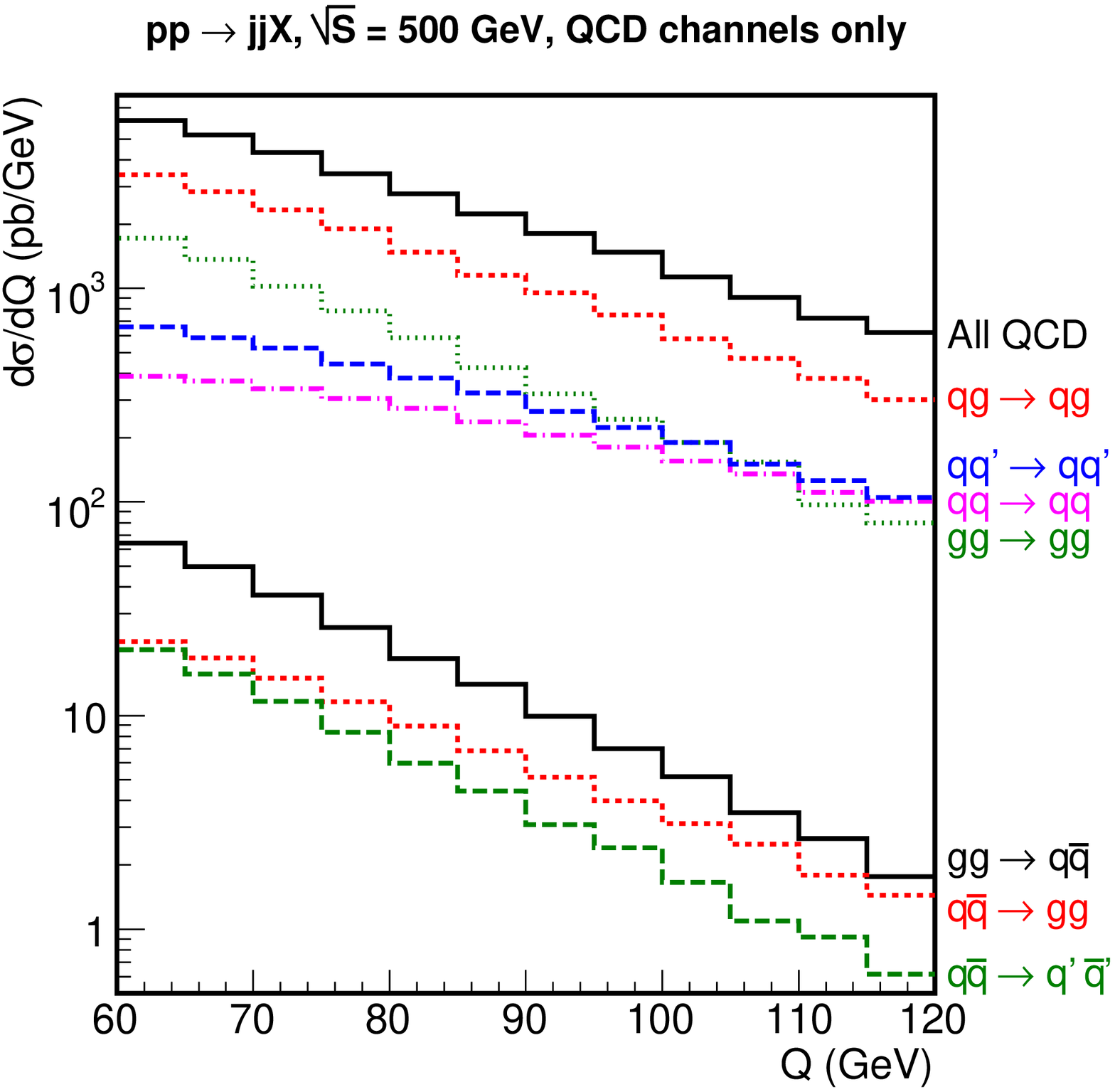}\includegraphics[width=0.5\columnwidth,keepaspectratio]{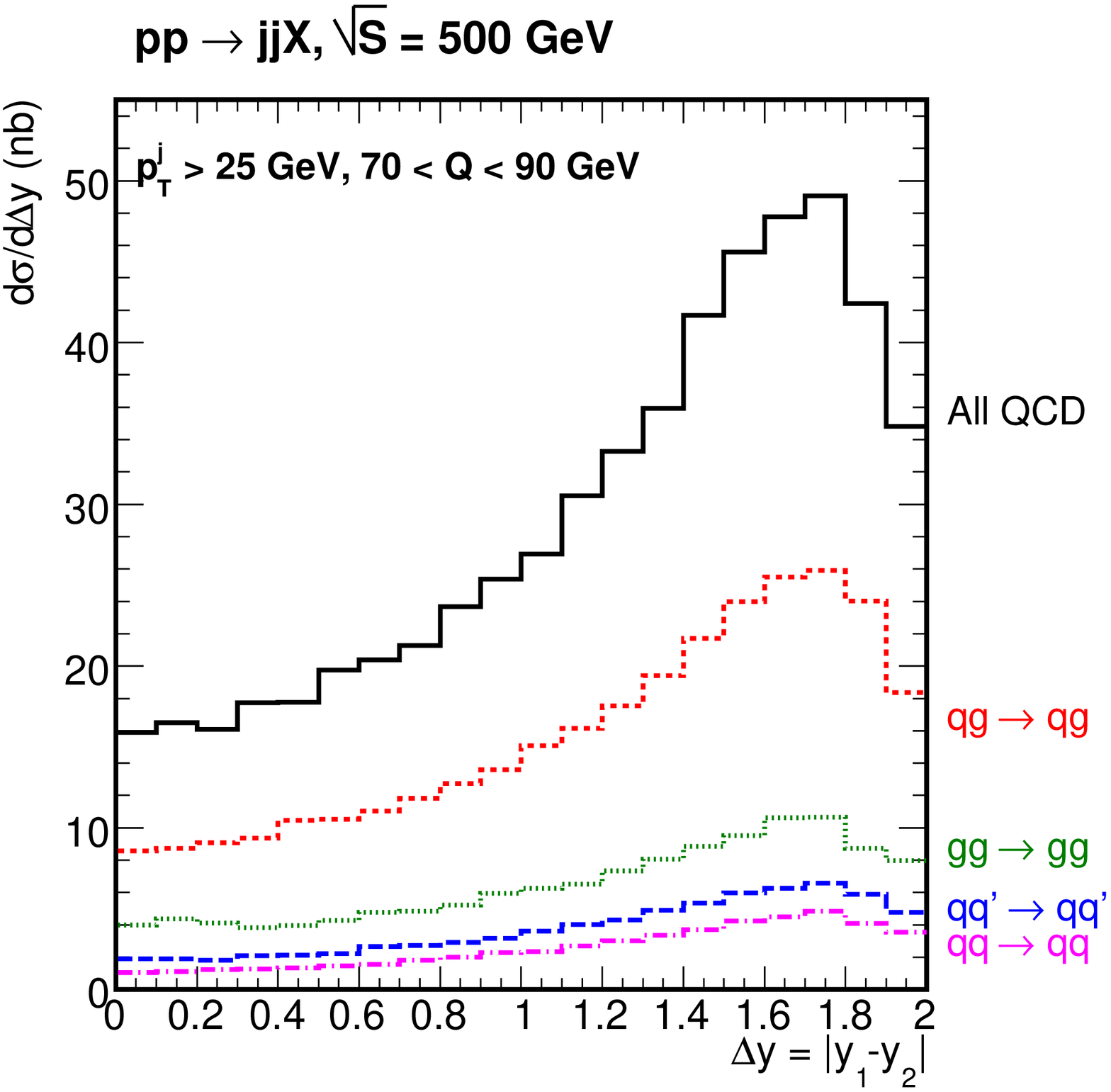}\\
 (a)\hspace{2in}(b)

\caption{${\cal O}(\alpha_{s}^{2})$ cross sections for $pp\rightarrow jjX$
at $\sqrt{s}=500$ GeV in various scattering channels (QCD contributions
only), plotted versus (a) dijet invariant mass $Q$, (b) difference
of jet rapidities $\Delta y=\left|y_{1}-y_{2}\right|$. \label{Fig:flavor_composition}}
\end{figure}

In Fig.~\ref{Fig:flavor_composition} we examine the flavor composition
and angular dependence of QCD scattering channels (proportional to
$\alpha_{s}^{2}$). In the relevant region of $Q,$ QCD jet production
is dominated by quark-gluon scattering, $qg\rightarrow qg,$ and,
at lower $Q$ values (not shown) by gluon-gluon scattering $gg\rightarrow gg$
(cf. Fig.~\ref{Fig:flavor_composition}(a)). Both types of contributions
are peaked strongly at large $\left|\Delta y\right|$; see Fig.~\ref{Fig:flavor_composition}(b).
The quark-quark scattering QCD contributions, $qq\rightarrow qq$
and $qq^{\prime}\rightarrow qq^{\prime}$, have a flatter dependence
on $Q$ and $\left|\Delta y\right|$ than the gluon-scattering contributions.
Similarly, the electroweak ${\cal O}(\alpha_{EW}^{2}+\alpha_{s}\alpha_{EW})$
dijet production, also proceeding via quark-quark scattering, has
a flatter dependence on $\left|\Delta y\right|$ than the gluon-scattering-dominated
QCD background.

\begin{figure}
\begin{centering}\includegraphics[width=8cm]{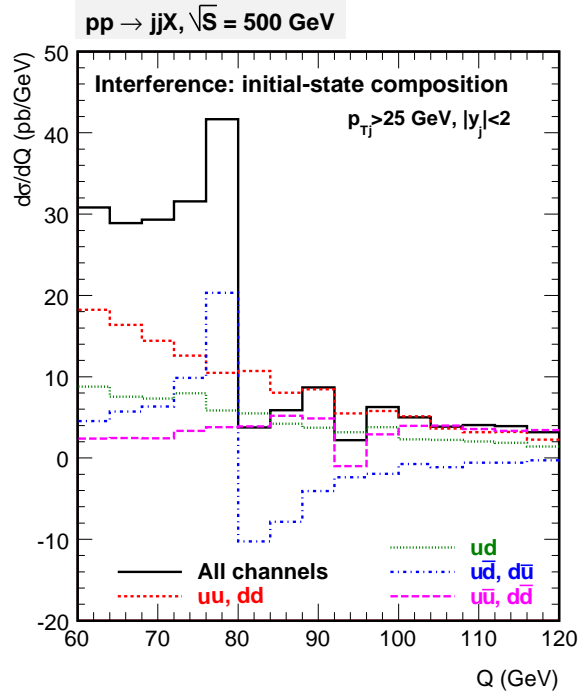} \par\end{centering}

\caption{The unpolarized interference cross section (solid) and contributions
of individual initial-state channels: $uu$ and $dd$ (short-dashed);
$ud$ (dotted); $u\bar{d}$ and $d\bar{u}$ (dot-dashed); $u\bar{u}$
and $d\bar{d}$ (long-dashed). \label{Fig:interf_composition}}
\end{figure}

The $qq\rightarrow qq$ and $qq^{\prime}\rightarrow qq^{\prime}$
QCD amplitudes can interfere with the electroweak amplitudes having
the same initial and final quark states. The largest contributions
to the interference, together with the full interference term, are
shown in Fig.~\ref{Fig:interf_composition} as a function of $Q$.
The magnitude of the ${\cal O}(\alpha_{s}\alpha_{EW})$ interference
is similar to that of the ${\cal O}(\alpha_{EW}^{2})$ cross section.
Its shape is determined by the interplay of the partial interference
terms.

We classify these partial terms according to their initial-state quark
composition. That is, the cross section {}``$uu,\, dd$'' corresponds
to the sum of the interference cross sections with two initial-state
$u$ quarks or two $d$ quarks, i.e., $uu\rightarrow uu$ and $dd\rightarrow dd$;
{}``$ud$'' corresponds to the interference cross section for $ud\rightarrow ud$,
$ud\rightarrow du$, $du\rightarrow ud$, and $du\rightarrow du$;
and so on. Smaller interference terms involving strange or charm quarks
are included in the total interference, but not shown separately.

It is useful to distinguish between the interference terms that possess
resonant properties and those that do not. The resonant interference
occurs between an $s$-channel $W$ or $Z$ boson amplitude and a
different conjugate amplitude with the same external states. The real
part of the heavy boson propagator in the $s$-channel amplitude changes
sign at the boson's mass pole, $Q=M_{V}$. Consequently a resonant
interference term exhibits a pronounced enhancement slightly below
the $W$ or $Z$ pole and a comparable enhancement of the opposite
sign immediately above the pole. In our case, the {}``$u\bar{d},$
$d\bar{u}$'' cross section (dot-dashed curve) is the most prominent
resonant term dominated by the interference with the $s-$channel
$W$ boson amplitude. It changes sign at $Q=M_{W}$. Despite its large
magnitude, its contribution to the integrated event rate in the signal
region $70<Q<90$ GeV is small, due to the cancellation between the
contributions of the opposite sign from below and above $Q=M_{W}$.
Similarly, a resonance driven by the $s$-channel $Z$ boson amplitude
occurs in the {}``$u\bar{u},$ $d\bar{d}$'' cross section (long-dashed
line) at $Q=M_{Z}$.

The largest interference contributions to the integrated rate in the
signal region arise from non-resonant $uu,$ $ud,$ and $dd$ processes
(short-dashed and dotted lines). These cross sections do not contain
a large resonant propagator, but rather are enhanced by a product
of two large valence-dominated quark PDFs, $u(x_{1},Q)u(x_{2},Q),$
etc. From this discussion, we conclude that the $u\bar{d}$ and $d\overline{u}$
interference contributions cancel when integrated over the signal
region. The largest surviving interference is due to the $uu,$ $ud,$
and $dd$ terms. As a result the interference strongly affects dependence
on $u$ and $d$ quark PDFs, but only marginally on the $\overline{d}$
antiquark PDF.

\subsection{Parton flavor composition of $d\sigma/dy$ \label{sub:Parton-flavor-composition}}

\begin{figure}
\begin{centering}\includegraphics[width=0.5\columnwidth]{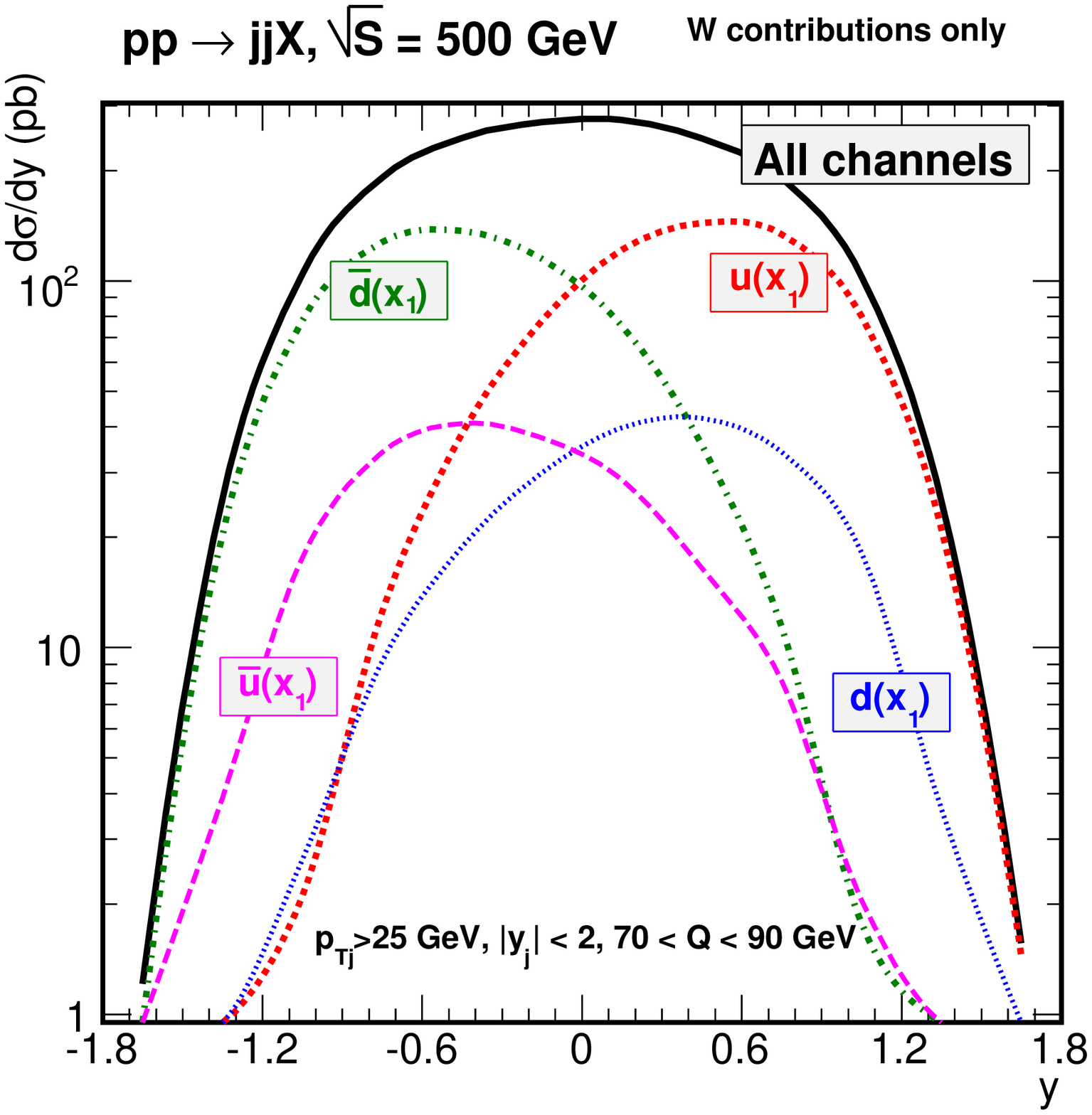}\includegraphics[width=0.5\columnwidth]{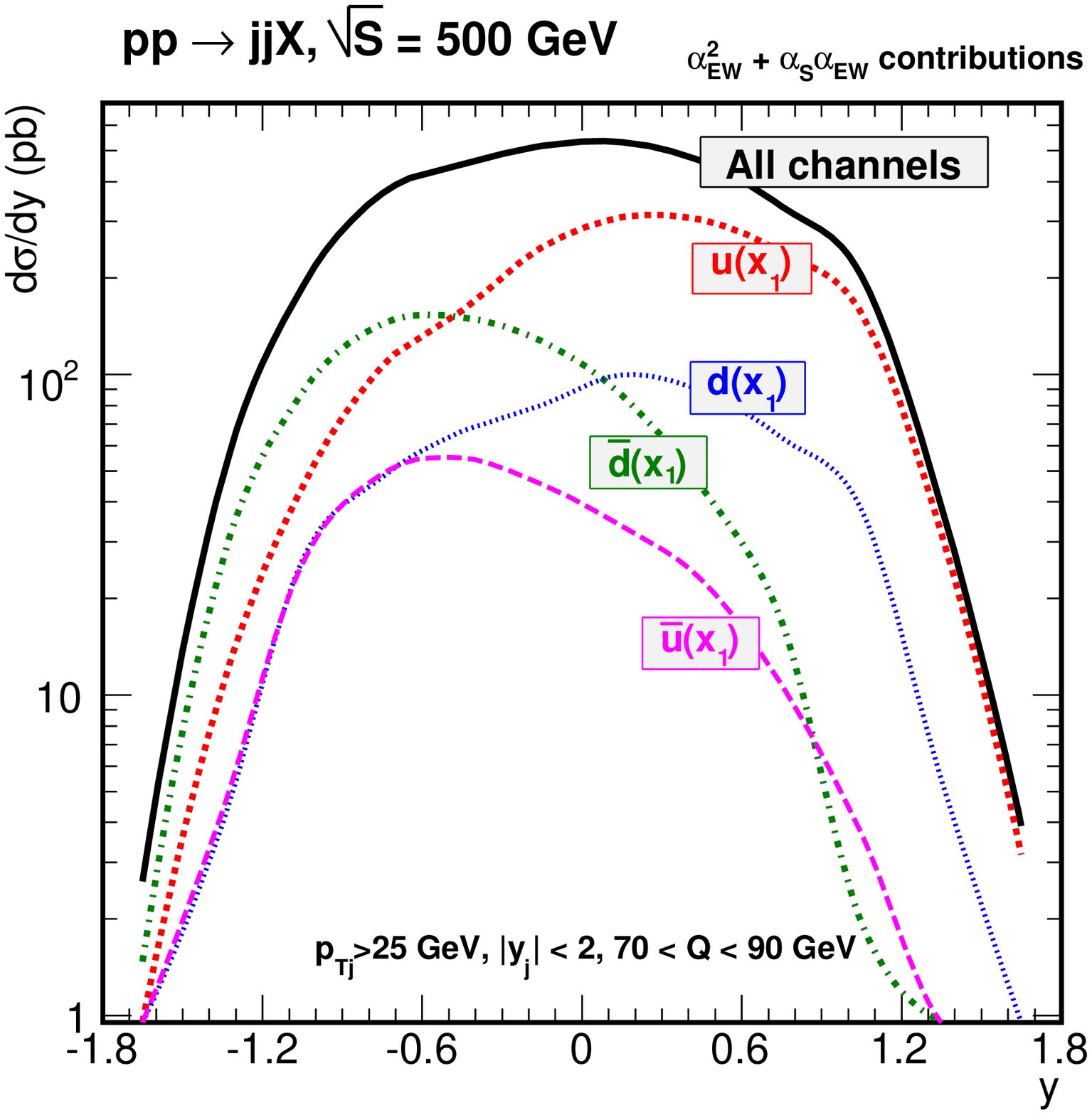}\\
 (a)\hspace{2.5in}(b)\par\end{centering}

\caption{Quark flavor composition of the unpolarized cross section $d\sigma/dy$:
(a) $W(\alpha_{EW}^{2})$ cross section only; (b) full electroweak+interference
${\cal O}(\alpha_{EW}^{2}+\alpha_{s}\alpha_{EW})$ cross section \label{fig:QuarkFlavor}}
\end{figure}

In Fig.~\ref{fig:QuarkFlavor}(a), we show partial contributions
to the ${\cal O}\left(\alpha_{EW}^{2}\right)$ cross sections $d\sigma/dy$
at $70<Q<90$ GeV involving only $W^{+}$ and $W^{-}$ contributions
in all channels. In this figure, we identify contributions proportional
to $u(x_{1}),$ $d(x_{1}),$ $\overline{u}(x_{1}),$ and $\bar{d}(x_{1})$.
The pure $W^{\pm}$ contribution is dominated by the resonant terms.
Hence the curves in Fig.~\ref{fig:QuarkFlavor}(a) closely follow
the rapidity dependence suggested by Eqs.~(\ref{sWplus}) and (\ref{sWminus}).
Only one term {[}proportional to $u(x)$ or $d(x)$ at $x\rightarrow1$]
survives on the right-hand sides of these equations when $y$ approaches
its kinematical limits, $y\rightarrow\pm\ln(Q/\sqrt{s})$. 

The cross section supplied by $W^{-}$ contributions constitutes about
1/3 of that from $W^{+}$ contributions. Consequently, the combined
$W^{\pm}$ cross section is dominated by $u(x_{1})$ and $\bar{d}(x_{1})$
contributions at large positive and negative $y$, as in resonant
$W^{+}$ production.

Figure~\ref{fig:QuarkFlavor}(b) shows the quark flavor decomposition
for the electroweak+interference ${\cal O}(\alpha_{EW}^{2}+\alpha_{s}\alpha_{EW})$
cross section. In this case, the combined $u(x_{1})$ contribution
is roughly equal to the sum of the $W^{+}$ contribution in Fig.~\ref{fig:QuarkFlavor}(a)
and $uu,$ $ud$ interference contribution (symmetric with respect
to $y=0$). The combined $d(x_{1})$ contribution is similarly made
of the $W^{-}$ and $dd,$ $du$ interference contributions. Both
$\overline{u}(x_{1})$ and $\overline{d}(x_{1})$ contributions remain
dominated by the resonant $W^{\pm}$ terms, as the interference involving
sea quarks is quite small. The essential conclusion to be drawn is
that the combined electroweak+interference cross section largely preserves
the PDF dependence of the resonant $W^{+}$ production, notably, the
sensitivity to $u(x_{1},Q)$ ($\overline{d}(x_{1},Q)$) at the forward
(backward) $y$ values.

\section{The single-spin asymmetry \label{sec:Asymmetry}}

\begin{figure}
\includegraphics[width=0.5\columnwidth,keepaspectratio]{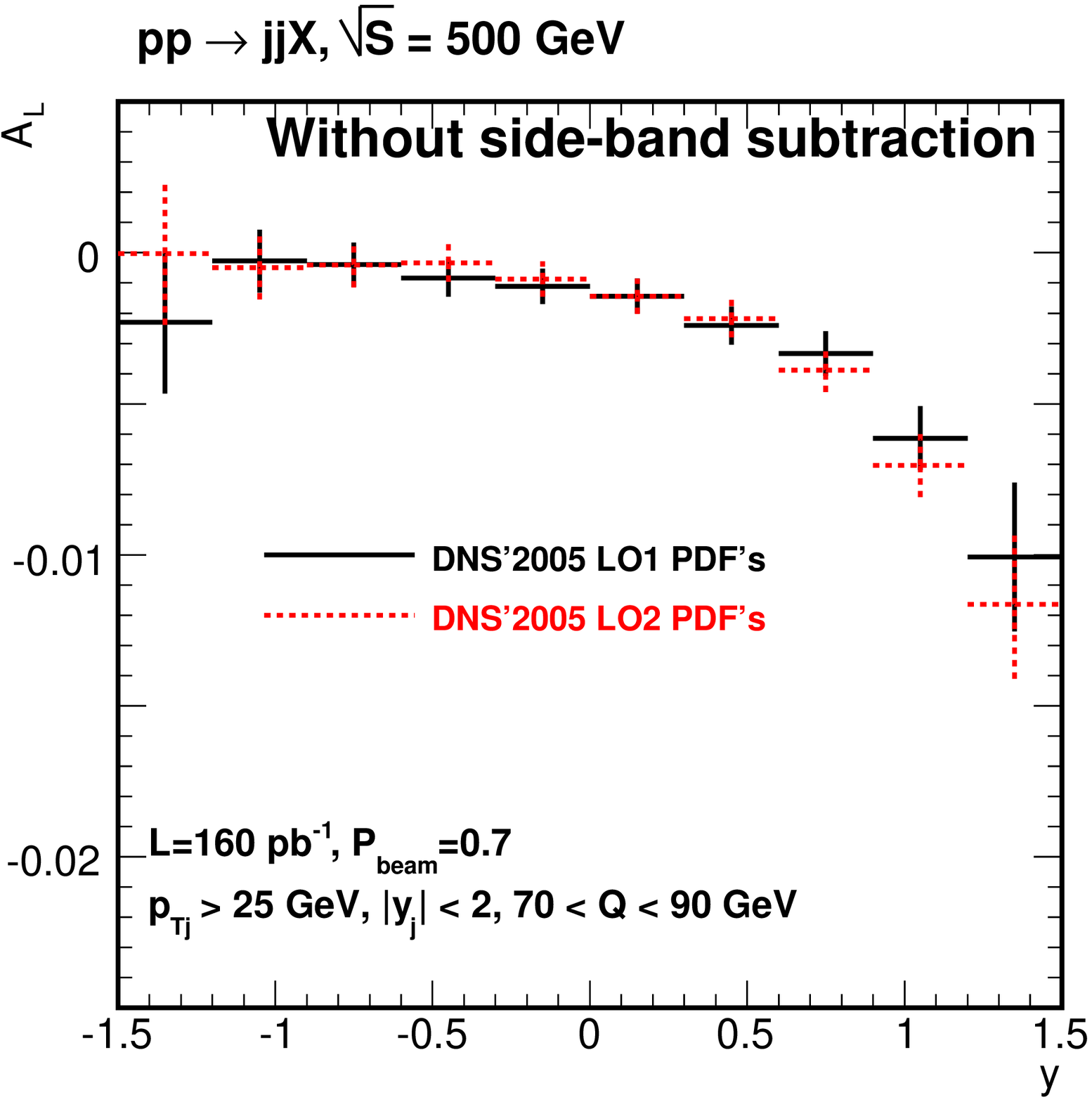}\includegraphics[width=0.5\columnwidth,keepaspectratio]{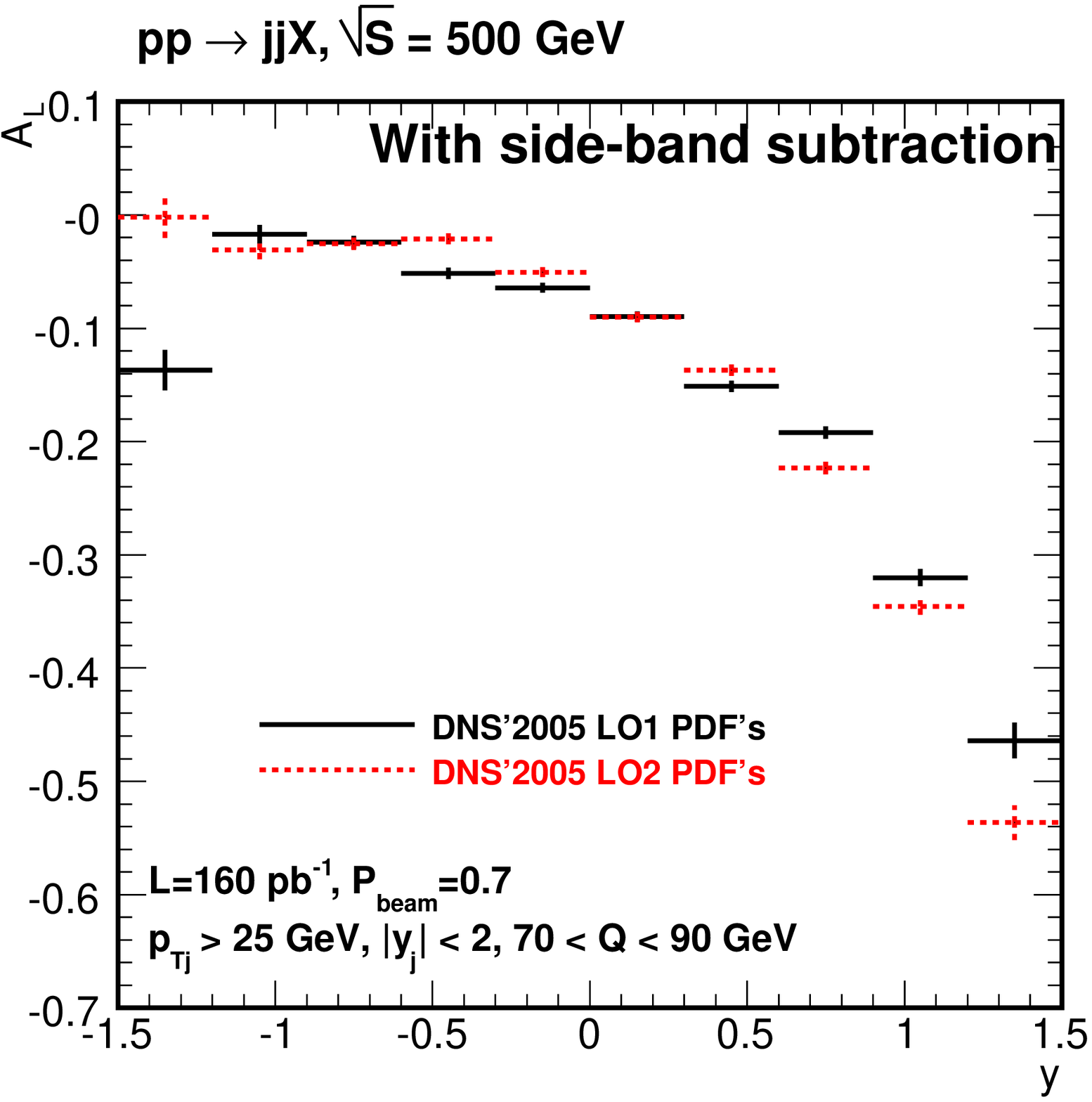}

\caption{Parity-violating asymmetry $A_{L}$ in the dijet mass interval $70<Q<90$
GeV without and with side-band subtraction of the parity-conserving
QCD background, plotted for two sets of leading-order DNS'2005 polarized
PDFs. \label{Fig:AL}}
\end{figure}

In Fig.~\ref{Fig:AL} we present the single-spin asymmetry $A_{L}(y)$,
evaluated as a function of rapidity for two sets of DNS'2005 leading-order
polarized PDFs and MRST'2002 NLO unpolarized PDFs. The statistical
uncertainty $\delta A_{L}$ in the measurement of the asymmetry is
evaluated as \cite{Bunce:2000uv} \begin{equation}
\delta A_{L}=\sqrt{\frac{1}{{\cal L}\sigma}\left(\frac{1}{P_{beam}^{2}}-A_{L}^{2}\right)}\label{deltaAL}\end{equation}
 for the integrated luminosity ${\cal {\cal L}}=160\mbox{ pb}^{-1}$
and proton beam polarization $P_{beam}=0.7$.

The asymmetry $A_{L}$ and its projected statistical uncertainty can
be determined in two ways. In the first (most straightforward) method,
the spin-averaged cross section is taken to be the full ${\cal O}\left(\alpha_{S}^{2}+\alpha_{s}\alpha_{EW}+\alpha_{EW}^{2}\right)$
cross section. 
The resulting rapidity ($y$) dependence of $A_{L}$ in the signal
region $70<Q<90$ GeV is shown in Fig.~\ref{Fig:AL}(a). The value
of $A_{L}$ is generally small, and the statistical uncertainties
(evaluated based on the total number of events for both beam polarizations
in each bin) do not allow discrimination between the sets of polarized
PDFs shown.

In the second method, we set the unpolarized cross section equal to
the ${\cal O}\left(\alpha_{s}\alpha_{EW}+\alpha_{EW}^{2}\right)$
cross section both in the denominator of $A_{L}$ and $\delta A_{L}.$
This procedure is roughly equivalent to the measurement of $A_{L}$
in which the background cross section (dominated by the ${\cal O}(\alpha_{s}^{2})$
term) is measured precisely in the side bands ($Q<70$ GeV and $Q>90$
GeV), extrapolated into the signal region ($70<Q<90$ GeV), and subtracted
from the measured unpolarized rate. Using this approximate side-band
subtraction technique in Fig.~\ref{Fig:AL}(b), we see that the predicted
magnitude of $A_{L}$ is increased substantially, and different parametrizations
of the polarized PDFs can be discriminated, given the projected statistical
uncertainties.

The asymmetry $A_{L}$ includes both the $W^{+}$ and $W^{-}$ contributions,
as a result of summation over all jet charges. At large positive $y$,
$A_{L}$ approximately follows $-\Delta u(x)/u(x)$ at large $x$,
expected to satisfy $\Delta u(x)/u(x)\rightarrow1$ in the $x\rightarrow1$
limit. At large negative $y,$ $A_{L}$ mostly reflects the behavior
of sea quark PDFs at $x\rightarrow0.02,$ notably of $\Delta\overline{d}(x)/\overline{d}(x).$

\begin{figure}
\includegraphics[width=0.6\columnwidth]{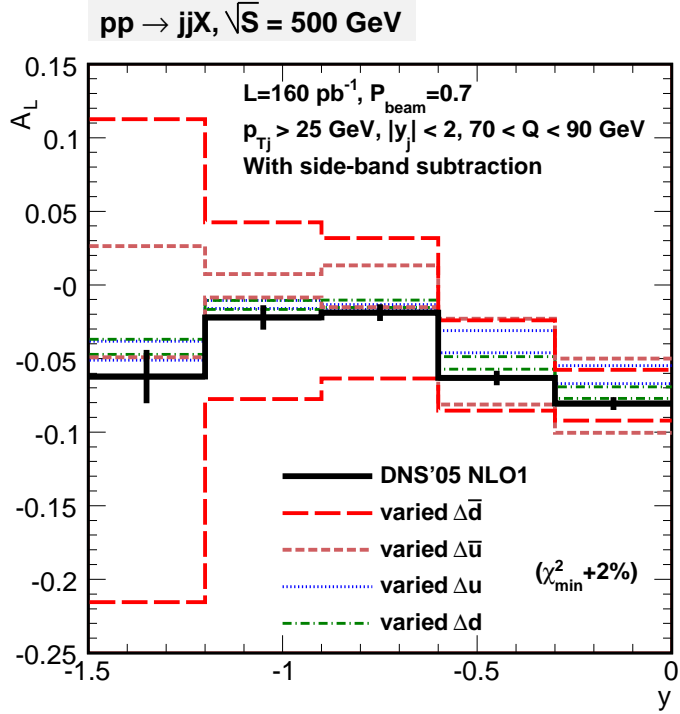}

\caption{The asymmetry $A_{L}$ for backward production of jet pairs ($y<0$),
computed for DNS'2005 NLO polarized PDFs \cite{deFlorian:2005mw}
after side-band subtraction. The black solid curve corresponds to
the best-fit PDF set 1. The pairs of other curves correspond to the
maximal and minimal values of $A_{L}$ obtained if the first moment
$\Delta q\equiv\int_{0}^{1}dx\Delta q(x,Q=3.16\mbox{ GeV})$ deviates
from its best-fit value by an amount corresponding to $\chi^{2}=1.02\chi_{min}^{2}$,
where $\chi_{min}^{2}$ is the minimum of the log-likelihood function
$\chi^{2}$ in the fit. We show $A_{L}$ for varied $\Delta\overline{d}$
(red long dashes); varied $\Delta\bar{u}$ (brown short dashes); varied
$\Delta u$ (blue dots); and varied $\Delta d$ (green dot-dashes).
At large negative rapidities, the most pronounced variations in $A_{L}$
are due to the variation of $\Delta\bar{d}.$ \label{fig:ALnegY}}
\end{figure}

We now examine the region of negative $y$ in greater detail in order
to gauge the sensitivity of $A_{L}$ to variations in the polarized
PDFs within the limits tolerated by current parametrizations. We plot
the values of $A_{L}$ obtained after side-band subtraction for the
DNS'05 NLO PDF sets with varied first moments of $\Delta u(x),$ $\Delta d(x),$
$\Delta\overline{u}(x),$ and $\Delta\overline{d}(x)$. The results
are shown in Fig.~\ref{fig:ALnegY}. When $y<-1$, $A_{L}$ is clearly
sensitive to the variation in the first moment of $\Delta\overline{d}$
and, to a smaller degree, of $\Delta\overline{u}$. It is nearly insensitive
to the variations in $\Delta u$ and $\Delta d$. This result suggests
that the measurement of $A_{L}$ will constrain $\Delta\overline{d}(x)$,
given the projected statistical uncertainties.

We remark that the pronounced sensitivity of $A_{L}$ to $-\Delta u(x)/u(x)$
and $\Delta\overline{d}(x)/\overline{d}(x)$ arises because of the
dominance of the $W^{+}$ contribution to parity-violating dijet production,
nearly independently of the PDF set chosen. Equations~(\ref{AL},
\ref{sWplus}-\ref{DsWminus}) state that the resonant parity-violating
part of $A_{L}$ behaves at large $\left|y\right|$ approximately
as \begin{equation}
\lim_{y\rightarrow-\ln\left(M_{W}/\sqrt{s}\right)}A_{L}(y)\approx-\frac{\frac{\Delta u(x_{1})}{u(x_{1})}+\frac{\Delta d(x_{1})}{d(x_{1})}r(x_{1},x_{2})}{1+r(x_{1},x_{2})},\label{ALlimx1}\end{equation}
and \begin{equation}
\lim_{y\rightarrow\ln\left(M_{W}/\sqrt{s}\right)}A_{L}(y)\approx\frac{\frac{\Delta\overline{d}(x_{1})}{\overline{d}(x_{1})}+\frac{\Delta\overline{u}(x_{1})}{\overline{u}(x_{1})}r(x_{2},x_{1})}{1+r(x_{2},x_{1})},\label{ALlimx2}\end{equation}
 where \begin{equation}
r(x_{1},x_{2})\equiv\frac{d(x_{1})}{u(x_{1})}\frac{\bar{u}(x_{2})}{\bar{d}(x_{2})}.\end{equation}
The factor $r$ is small (of order 1/3) in both limits, as a consequence
of the smallness of the $d(x)/u(x)$ ratio at $x\rightarrow1$ (see,
e.g., Figs.~3-6 in Ref.~\cite{Owens:2007kp}). Unless $\Delta u(x_{1})/u(x_{1})$
is much smaller than $\Delta d(x_{1})/d(x_{1})$ in absolute magnitude,
it dominates the numerator of $A_{L}(y)$ in Eq.~(\ref{ALlimx1}).
Similarly, $\Delta\bar{d}(x_{1})/\overline{d}(x_{1})$ dominates the
numerator of Eq.~(\ref{ALlimx2}), unless it is much smaller than
$\Delta\overline{u}(x_{1})/\overline{u}(x_{1})$ in absolute magnitude. 

In the lepton decay mode, the positron from a $W^{+}$ decay tends
to scatter into the central rapidity region, hence smearing the $x$
dependence of the underlying $\overline{d}$ parton distribution \cite{Nadolsky:2003ga,RHICSpin2008}.
In jet production, the smearing does not occur as a result of the
direct measurement of the rapidity $y$ of the jet pair.

Observation of $A_{L}$ in the jet pair mode, after the background
subtraction, appears to guarantee a large asymmetry associated with
$\Delta u(x)$ at $x\rightarrow1$ for forward jet pair rapidities
and to test $\Delta\overline{d}(x)$ in the region of negative jet
pair rapidities. %

\section{Conclusions and Discussion \label{sec:Conclusions}}

Data on $W$ production obtained from longitudinally polarized proton-proton
scattering at RHIC can be used to extract spin-dependent quark and
antiquark parton distributions $\Delta q_{i}(x,M_{W})$ and $\Delta\bar{q}_{i}(x,M_{W})$,
providing information complementary to that obtained from polarized
deep-inelastic lepton scattering. The longitudinal spin asymmetry
$A_{L}(y)$ of the rapidity ($y$) dependence of $W$ production is
particularly sensitive to the spin-dependent quark PDFs. In this paper,
we examine the prospects for definitive measurements of $A_{L}(y)$
when the $W$ boson is detected at RHIC in its hadronic decay mode,
$W\rightarrow2$~jets, at the collision energy $\sqrt{s}=500$~GeV.

To take advantage of the increase in the event rate and, in principle,
of the direct measurement of $y$ offered by the hadronic decay mode,
one must first address the challenges of the large parity-conserving
background from QCD and electromagnetic production of jets in the
vicinity of the $W$ signal. In Sec.~\ref{sec:Unpolarized} of this
paper, we report our tree-level computation of the amplitudes for
all electroweak and strong interactions processes that lead to a pair
of jets in proton-proton scattering. We treat carefully all effects
of interference among the subprocesses. As shown in Fig.~\ref{Fig:unp},
the backgrounds in unpolarized scattering appear admittedly daunting.
Most of this background is due to subprocesses with emission of one
or two final-state gluons, as shown in Fig.~\ref{Fig:flavor_composition}.
Nevertheless, even at the unpolarized level, an experimental side-band
subtraction procedure could be used to reach acceptable sensitivity
to events containing two final-state quarks from electroweak scattering
or QCD-electroweak interference. With this procedure in place, the
unpolarized event rate is dominated by processes involving the $u$
quark PDF in the forward region of the dijet $y$, and by processes
involving the $\bar{d}$ antiquark PDF in the backward region of $y$,
even without separation of dijets from $W^{+}$ and $W^{-}$ production
{[}cf.~Fig.~\ref{fig:QuarkFlavor}(b)].

Turning to the longitudinally polarized case, we remark first that
the largest QCD and electromagnetic backgrounds cancel in the parity-violating
numerator $\sigma(p^{\rightarrow}p)-\sigma(p^{\leftarrow}p)$ of $A_{L}(y)$.
This means that the desired sensitivity of $A_{L}(y)$ to $\Delta u(x,M_{W})$
and $\Delta\overline{d}(x,M_{W})$ is mostly preserved. The magnitude
of $A_{L}(y)$ can be enhanced by applying the above subtraction procedure
to the spin-averaged denominator $\sigma(p^{\rightarrow}p)+\sigma(p^{\leftarrow}p)$
of $A_{L}(y).$ The strong and electromagnetic backgrounds may be
measured in the dijet invariant mass distribution in side-band regions
next to the $W$ signal location. An appropriate extrapolation and
subtraction of the backgrounds may then be performed when defining
the denominator of $A_{L}(y)$.

Our quantitative predictions for the asymmetry are presented in Figs.
\ref{Fig:AL} and \ref{fig:ALnegY}. In Fig.~\ref{Fig:AL}, we demonstrate
that, once the side-band subtraction technique is employed, the predicted
magnitude of $A_{L}$ is sufficiently large, when compared to the
expected uncertainties. The data on $A_{L}(y)$ would thus be a very
valuable component of a global analysis leading to definitive longitudinally
polarized PDFs.

At large positive $y$, $A_{L}$ approximately tracks $-\Delta u(x)/u(x)$
at large $x$, commonly assumed to satisfy $\Delta u(x)/u(x)\rightarrow1$
at $x\rightarrow1$. We find that $A_{L}$ can drop as low as $-0.5$
at $y=1.2-1.5$ (Fig.~\ref{Fig:AL}(b)). The statistical error in
this range of $y$ is still small enough to allow a meaningful measurement.

At large negative $y,$ $A_{L}$ reflects the behavior of the sea
quark density $\Delta\overline{d}(x)/\overline{d}(x).$ One way to
estimate the power of discrimination is to compute the changes in
the predicted $A_{L}(y)$ that result from varying the polarized parton
densities within the bands of tolerance of current parametrizations.
As shown in Fig.~\ref{fig:ALnegY}, the most pronounced predicted
variations in the backward rapidity region result from changes in
$\Delta\bar{d}$.

In an effort to increase the magnitude of $A_{L}(y)$, several techniques
may prove to be helpful. Selection of events at central $\Delta y=\left|y_{1}-y_{2}\right|$
(or central $\cos\theta_{*}$ in the dijet rest frame) helps to suppress
the gluon-dominated QCD background, as follows from Figs.~\ref{Fig:unp}(c)
and \ref{Fig:flavor_composition}(b).

One may consider selecting jets containing a final-state charmed particle
\cite{Arnold:2008zx}. In this case, the QCD background is reduced
much more than the $W$ signal rate by excluding many sources of light-flavored
jets. Using realistic efficiencies for charm detection, we cannot
conclude that the charm tag clearly improves the expected significance
of the measurements of $A_{L}(y)$. An alternative approach could
involve a tag on a leading charged particle in the jet (e.g., a charged
pion carrying at least a third of the total jet energy) to emphasize
contributions of the quark-initiated jets over gluon-initiated jets.
As with the charm tags, the reduction in the background over signal
in this case must be balanced against the substantial suppression
in the total event rate. We therefore defer our conclusion about the
feasibility of charm and leading-particle tags until a more detailed
study.

The analysis reported here uses hard-scattering cross sections computed
at the lowest order in perturbation theory. Next-to-leading order
(NLO) contributions in QCD increase the predicted rates for hadronic
jet production~\cite{Ellis:1994dg,Giele:1994xd} and are essential
for reducing large scale dependence of the lowest-order cross section.
This increase means that the backgrounds in the denominator of $A_{L}(y)$
can be larger than our estimates. On the other hand, as shown by Moretti
{\em et al.}~\cite{Moretti:2005aa}, comparably large enhancements
are predicted in the single spin polarized cross sections at RHIC,
meaning naively that the predicted magnitude of $A_{L}(y)$ could
remain largely unaffected. A quantitative statement about the ultimate
effects of NLO terms on $A_{L}(y)$ would require a consistent treatment
of NLO contributions to all contributing processes, in both the unpolarized
and polarized cases, as well as consistent application of kinematic
selections that we find advantageous. We leave this task for future
work, but we note here that comparably large enhancements of both
the polarized and unpolarized rates would lead to a {\em decrease}
in the projected statistical uncertainty $\delta A_{L}$, Eq.~(\ref{deltaAL}),
since $\delta A_{L}$ is proportional to $1/\sqrt{N}$, where $N$
is the number of events observed.

Our essential conclusion is that an experimental study of high-$p_{T}$
jet pair production at RHIC appears to offer a promising way to make
definitive measurements of the longitudinal spin asymmetry $A_{L}(y)$.

\begin{acknowledgments}
The motivation for this work and preliminary estimates were discussed
by P.~M.~N. at RHIC Spin Workshops at Brookhaven National Laboratory
in October, 2005 and April 2007. We are grateful to the organizers
of these workshops for the stimulating environment. P\@.~M.~N.
is particularly grateful to S. Arnold, A. Metz, and W. Vogelsang for
enlightening discussions of the presented topics.

E.\ L.\ B.\ is supported by the U.~S.\ Department of Energy under
Contract No.\
DE-AC02-06CH11357. P.~M.~N. is partly supported by the U.S.\ Department
of Energy under grant DE-FG02-04ER41299, and by Lightner-Sams Foundation.
The authors thank the Kavli Institute for Theoretical Physics (KITP),
Santa Barbara, for hospitality during the course of some of this work.
The KITP is supported by the National Science Foundation under Grant
No.\ NSF PHY05-51164. 
\end{acknowledgments}


\end{document}